\documentclass[letterpaper,journal]{IEEEtran}
\usepackage{cite}
\usepackage{float}
\usepackage{amsmath,amssymb,amsfonts}
\usepackage{algorithmic}
\usepackage{graphicx}
\usepackage{textcomp}
\usepackage{hyperref}
\def\BibTeX{{\rm B\kern-.05em{\sc i\kern-.025em b}\kern-.08em
    T\kern-.1667em\lower.7ex\hbox{E}\kern-.125emX}}
\usepackage{mathtools}
\usepackage{algorithmic}
\usepackage{algorithm}
\usepackage{array}
\usepackage[caption=false,font=normalsize,labelfont=sf,textfont=sf]{subfig}
\usepackage{textcomp}
\usepackage{stfloats}
\usepackage{booktabs} 
\usepackage{adjustbox, diagbox, adjustbox}
\usepackage{tabularx, multirow, multicol, makecell}
\begin{document}

\title{Feasibility of Time-Domain DNN-Based Speech Enhancement on Embedded FPGA for Hearing Aids}

\author{Feyisayo Olalere,~Umut Altin,~Kiki van der Heijden,~and Marcel van Gerven%
\thanks{F. Olalere and U. Altin contributed equally to this work.}%
\thanks{F. Olalere, U. Altin, and M. van Gerven are with the Donders Institute 
for Brain, Cognition, and Behaviour, Radboud University, Nijmegen, 
The Netherlands (e-mail: \texttt{feyisayo.olalere@donders.ru.nl}; 
\texttt{umut.altin@donders.ru.nl}; \texttt{marcel.vangerven@donders.ru.nl}).}%
\thanks{K. van der Heijden is with the Donders Institute for Brain, Cognition, 
and Behaviour, Radboud University, Nijmegen, The Netherlands, and also with 
the Mortimer B. Zuckerman Mind, Brain, Behavior Institute, Columbia University, 
New York, NY, USA (e-mail: \texttt{kiki.vanderheijden@donders.ru.nl}).}%
\thanks{This work is part of the INTENSE consortium, which has received funding 
from the NWO Cross-over Grant No.~17619.}%
}

\maketitle

\begin{abstract}
Hearing aids impose strict latency and power constraints that current DNN-based speech enhancement systems struggle to meet on embedded hardware. We characterize this gap by deploying both speech separation and denoising using the lightweight SuDoRM-RF++ architecture on the AMD-Xilinx Kria KV260, evaluated at FP32 and 16-bit fixed-point precision for each task. Across these configurations, first-sample latency tracks with on-chip parameter caching rather than arithmetic throughput, identifying data movement as the primary bottleneck. Precision reduction halves the model memory footprint without compromising objective speech quality. The fixed-point denoising accelerator achieves a first-sample latency of 9.7~ms, meeting the 10~ms clinical threshold, while speech separation reaches 16.0~ms. These measurements establish concrete resource requirements for embedded DNN-based speech enhancement and quantify the remaining gap to hearing aid deployment.
\end{abstract}


\section{Introduction}
\label{sec:intro}

The ``cocktail party problem," defined as the ability to selectively attend to a target speaker amidst competing talkers and background noise, remains a fundamental challenge for users of hearing prosthetics~\cite{bronkhorst2015cocktail}. While earlier research explored digital signal processing techniques to amplify a sound source of interest and attenuate other sound sources in the scene, such approaches used in commercial hearing aids continue to struggle in complex acoustic environments. Traditional methods, such as beamforming, directional microphone arrays, and spectral subtraction, have demonstrated efficacy in stationary noise conditions where acoustic characteristics remain relatively constant~\cite{loizou2007speech, boll2003suppression}. However, these classical techniques rely on assumptions of noise stationarity and exhibit significant performance degradation in the non-stationary environments typical of everyday life, such as busy restaurants or multi-talker conversations where acoustic conditions fluctuate rapidly \cite{loizou2007speech}. Adaptive approaches, including Wiener filtering and minimum variance distortionless response (MVDR) beamformers, attempt to address temporal variations but require accurate noise statistics and often introduce audible artifacts during rapid transitions~\cite{ephraim2003speech}. Ultimately, the dependence on hand-crafted features and statistical assumptions fails to capture the complex, non-linear nature of the everyday acoustic environment. This leaves a performance gap in the environments where user assistance is most needed.

To overcome these limitations, recent advances in deep neural networks (DNNs) have improved speech enhancement capabilities. DNNs have proven effective for two complementary tasks: speech separation, where the goal is to isolate individual speech sources from multi-talker mixtures, and speech denoising, which aims to suppress background noise while preserving the target speaker's intelligibility. In speech separation, models process either spectral representations in the time-frequency domain~\cite{narayanan2013ideal,lu2013speech,sun2020supervised} or raw waveforms in the time domain~\cite{zhang2016deep, luo2019conv}. Time-domain architectures have gained prominence due to several advantages: they avoid phase-reconstruction artifacts that commonly degrade audio quality, exhibit lower computational complexity, and achieve lower algorithmic latency~\cite{luo2019conv, luo2020dual, tzinis2022compute}. Speech denoising models have similarly adopted time-domain processing~\cite{lu2022conditional,leglaive2020recurrent}, with the distinction that they output a single enhanced speech signal rather than multiple separated sources. Despite their performance improvements on single-channel audio, with state-of-the-art models achieving signal-to-distortion ratios exceeding 15 dB for separation ~\cite{luo2019conv,subakan2021attention} and perceptual quality scores approaching those of clean speech for both tasks ~\cite{defossez2020real,lu2022conditional}, these architectures remain computationally intensive. Current leading models typically contain millions of parameters and require GPU acceleration or multi-core CPUs to maintain real-time processing, consuming hundreds of milliwatts to several watts of power ~\cite{gerlach2022survey}. In contrast, hearing aids operate within a highly constrained power envelope, typically requiring the entire signal-processing chain to consume between 1 and 3 mW~\cite{kates2008hearing,sun2020supervised}. Furthermore, these devices must maintain a total processing latency of less than 10 ms to prevent the comb-filter effect, which is a spectral distortion caused by the interference of processed audio with natural sound leaking through the ear canal, and to ensure that the user's own voice remains perceptually natural~\cite{stone2003tolerable}.

Field-Programmable Gate Arrays (FPGAs) can serve as a platform for bridging the gap between high-performance DNNs and the hardware constraints of small devices such as hearing aids~\cite{guo2016model}. FPGAs as accelerators enable fine-grained parallelism, which can also help reduce the model's processing latency~\cite {shawahna2018fpga}. Embedded FPGA Systems-on-Chip (SoCs) in particular offer a resource-constrained but reconfigurable substrate well-suited for characterizing the feasibility of DNN deployment under conditions that approximate the tight power and memory budgets of hearing aids~\cite{ni2025fpga}. Additionally, FPGAs are more readily accessible and flexible for prototyping than specialized ASIC chips, which are more expensive and require extensive fabrication~\cite{park2024real, itani2025tf}. Modern FPGA development workflows leverage High-Level Synthesis (HLS) tools, which allow neural network inference pipelines to be described in C++ and automatically compiled into hardware logic, significantly reducing implementation time compared to manual register transfer level design. 

Earlier FPGA implementations of speech enhancement algorithms focused on mapping classical signal processing techniques, such as adaptive beamforming~\cite{enemali2023towards}, comb filters for binaural spectral splitting~\cite{kambalimath2014fpga}, and cochlear-mimicking filter channels~\cite{enemali2023towards, mishra2002cochlear, yang2015per}, directly onto reconfigurable hardware. These designs leveraged the inherent parallelism of FPGAs to implement customized structures, including FIR converters~\cite{rao2023design} and sequential multiply-accumulate (MAC) operations~\cite{kambalimath2014fpga}, to address spectral and temporal masking effects. By optimizing these classical algorithms, researchers demonstrated real-time performance and significantly lower power consumption on FPGAs compared to general-purpose processors.

However, as mentioned earlier, classical algorithms have their limitations in non-stationary interference or background noise. While deep learning offers better performance in these complex environments, few studies have systematically mapped DNN-based enhancement algorithms onto FPGAs due to memory constraints. This is the challenge of fitting millions of model parameters into the limited on-chip Block RAM (BRAM) of embedded devices without incurring the latency penalties of accessing external memory. Current research has largely focused on isolated solutions. In~\cite{ni2025fpga}, an encoder-decoder LSTM architecture was accelerated on an AMD RFSoC42 to suppress environmental noise, and while they improved the latency in comparison to SoC CPUs and Raspberry Pi, their real-time factor was still greater than one (RTF $>$ 1), which makes it less ideal for real-time use. Beyond throughput, first-sample latency must also meet the 10-ms clinical threshold \cite{stone2003tolerable}. Flamis et al.~\cite{flamis2022fpga} deployed a fully convolutional DNN on a Zynq-7000 SoC-FPGA using INT8 quantization, demonstrating DNN-based denoising on resource-constrained hardware, but neither latency nor RTF are reported, leaving real-time feasibility uncharacterized. Alternatively, a hybrid approach was proposed in~\cite{vasudev2025hybrid}, where a machine learning-based classifier is used to identify the acoustic environment and dynamically tune classical noise-reduction filters. While this modular strategy significantly reduces memory requirements by avoiding a full DNN-based enhancement architecture, the reliance on frequency-domain processing (such as spectral subtraction) introduces inherent algorithmic latency dictated by the Short-Time Fourier Transform (STFT) windowing and the step interval between consecutive analysis windows (hop-size). In contrast, time-domain implementations prioritize faster processing to meet clinical standards for hearing aids.

In this work, we present a systematic feasibility study of time-domain speech enhancement models deployed fully on an embedded FPGA platform, characterizing the gap between current embedded hardware capabilities and the latency constraints of hearing aids. Unlike prior work that focuses on single-task implementations or hybrid classical-DNN approaches, we implement both speech separation and speech denoising using a recent lightweight time-domain architecture~\cite{tzinis2022compute} on the Xilinx Kria KV260, a resource-constrained embedded SoC~\cite{xilinx2021kria}, evaluating both FP32 and F16 floating-point precision.  We address three questions: (1) What are the computational and memory requirements for deploying time-domain speech enhancement on an embedded FPGA? (2) How does embedded FPGA processing compare to CPU-based inference in terms of latency, and how does the FPGA platform hold up with power consumption? (3) Does F16 quantization provide meaningful latency gains while preserving acceptable audio quality? Our measurements show that the embedded FPGA achieves lower processing latency than a CPU baseline across both precision levels, and that F16 quantization provides a substantial latency reduction for both tasks, narrowing the gap toward the 10 ms clinical threshold. While current embedded FPGAs do not yet fully achieve all the constraints required for a hearing prosthetic, these results establish a concrete trajectory toward clinically viable deployment.

The remainder of this paper is structured as follows: Section II details our methodology, including the formal problem definition, dataset selection, time-domain model training, and the FPGA-specific deployment procedures for the KV260 platform. Section III presents the experimental results, characterizing the performance, power, and latency trade-offs for both speech separation and denoising tasks across FP32 and F16 precision. Section IV discusses the broader implications of these findings for AI-augmented hearing and the feasibility of DNN deployment in wearable form factors. Finally, Section V concludes the paper.

\section{Methodology}

\subsection{Problem Definition}

The goal of this study is to deploy single-channel speech enhancement models on FPGA platforms. We consider two complementary enhancement tasks: speech separation and speech denoising. 

In the speech separation task, the objective is to decompose a mixture of $N$ concurrent speakers into individual source signals. Given a single-channel mixture signal $\mathbf{y}(t) \in \mathbb{R}^T$, where $T$ represents the number of time samples, the mixture is formed as the sum of $N$ source signals:
\begin{equation}
    \mathbf{y}(t) = \sum_{i=1}^{N} \mathbf{s}_i(t)
\end{equation}
where $\mathbf{s}_i(t) \in \mathbb{R}^T$ denotes the $i$-th clean source signal. The separation model $f_{\text{sep}}(\cdot; \theta_{\text{sep}})$ parameterized by $\theta_{\text{sep}}$ estimates all $N$ source signals:
\begin{equation}
    \{\hat{\mathbf{s}}_1(t), \hat{\mathbf{s}}_2(t), \ldots, \hat{\mathbf{s}}_N(t)\} = f_{\text{sep}}(\mathbf{y}(t); \theta_{\text{sep}}),
\end{equation}
where $\hat{\mathbf{s}}_i(t) \in \mathbb{R}^T$ is the estimated signal for the $i$-th source. For this work, we focus on the two-speaker case ($N=2$). The model outputs two separated speech signals corresponding to the two speakers present in the mixture.

In the speech denoising task, the objective is to extract a single target speech signal from a noisy acoustic scene while suppressing background interference. Given a noisy input signal $\mathbf{y}(t) \in \mathbb{R}^T$:
\begin{equation}
    \mathbf{y}(t) = \mathbf{s}(t) + \mathbf{n}(t),
\end{equation}
where $\mathbf{s}(t) \in \mathbb{R}^T$ is the clean target speech and $\mathbf{n}(t) \in \mathbb{R}^T$ represents additive noise, the denoising model $f_{\text{den}}(\cdot; \theta_{\text{den}})$ parameterized by $\theta_{\text{den}}$ estimates the clean speech signal:
\begin{equation}
    \hat{\mathbf{s}}(t) = f_{\text{den}}(\mathbf{y}(t); \theta_{\text{den}}),
\end{equation}
where $\hat{\mathbf{s}}(t) \in \mathbb{R}^T$ is the enhanced output signal. Unlike speech separation, which produces multiple outputs, denoising produces a single enhanced signal that preserves the target speaker's voice while attenuating background noise sources such as environmental sounds, non-stationary noise, and reverberation.

\subsection{Datasets}
For the speech separation task, we used the standard WSJ0-2mix dataset, derived from the Wall Street Journal (WSJ0) corpus~\cite{hershey2016deep}. The dataset was partitioned into 20,000 training, 5000 validation, and 3000 test two-talker mixtures following the conventional splits established in the literature~\cite{huang2014deep,zhang2016deep,luo2019conv}. Each audio mixture was generated by combining two speakers at Signal-to-Noise Ratios (SNRs) ranging from -5 to 5 dB. Consistent with standard benchmarks for separation tasks, the audio was sampled at 8 kHz and cut to a duration of 4 seconds~\cite{tzinis2022compute,luo2019conv}. To prepare the data for the model, we applied mean-variance normalization to the input waveforms, ensuring a zero-mean and unit-variance distribution.

The speech denoising task was conducted using the Valentini-Botinhao dataset, which combines the Voice Bank corpus with background noise from the DEMAND database~\cite{valentini2017noisy,botinhao2016speech}. We use the standard 28-speaker training split containing 11,572 utterances ({\small$\sim$}28.2 hours) sampled at 16 kHz, with utterances ranging from 3.3 to 45.3 seconds in duration. The training set includes 10 noise types (2 artificial and 8 real-world recordings) mixed at four SNR levels (15, 10, 5, 0 dB). For evaluation, we use the standard test set with 2 speakers and 5 noise conditions not present in the training data. During training, we applied data augmentation from~\cite{defossez2020real}: random time shifts (0 to stride length), Remix augmentation (shuffling noise within batches), and BandMask (a band-stop filter removing 20\% of frequencies uniformly in the mel scale).  The BandMask augmentation is equivalent in the waveform domain to SpecAugment originally done in~\cite{park2019specaugment}. Prior to model input, we applied variance normalization by scaling the signal by its standard deviation.

\subsection{Time-Domain Model Architecture and Training}

\begin{figure*}[!ht]
    \centering
    \includegraphics[width=0.9\linewidth]{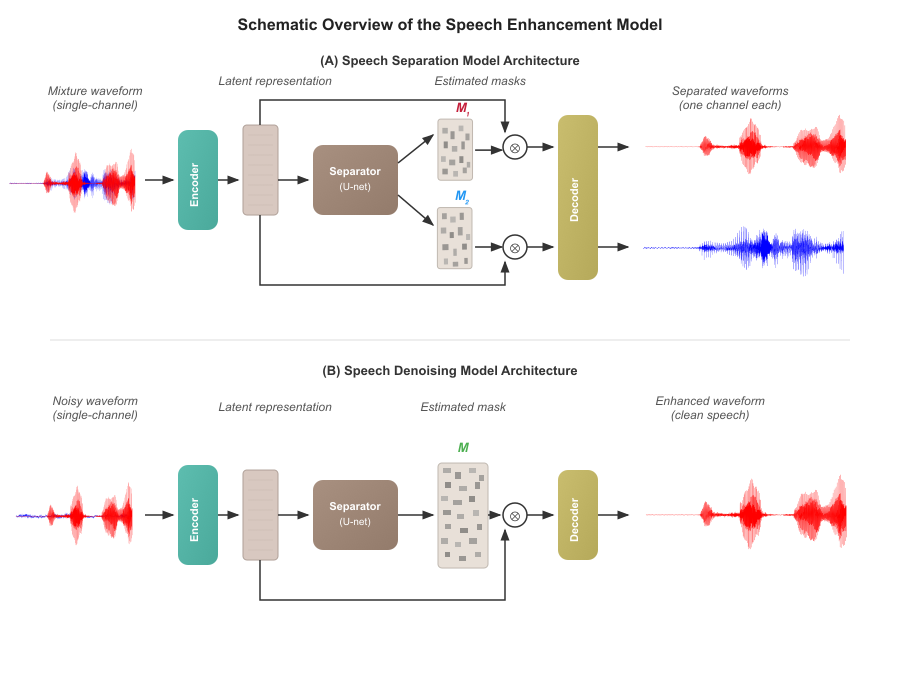}
    \caption{Schematic diagram for a single-channel speech enhancement algorithm~\cite{tzinis2022compute}. (A) Shows the original model as proposed for speech separation. (B) Shows the model adapted for speech denoising.}
    \label{img:AdaptedModel}
\end{figure*}

For this study, we utilize the causal SuDoRM-RF++ 0.25x architecture~\cite{tzinis2022compute}. This end-to-end time-domain model is designed for high-fidelity audio processing with minimal computational overhead by using Successive Downsampling and Resampling of Multi-Resolution Features. Rather than relying on expensive dilated convolutions, the architecture employs a series of U-ConvBlocks within its separator to capture long-term temporal dependencies through successive temporal resampling. This design ensures a compact memory footprint and low algorithmic latency, with an encoder kernel introducing only ~2.63 ms of delay, well below the minimum 32 ms latency of time-frequency based architectures ~\cite{isik2016single,kolbaek2017multitalker}. This low latency makes it well-suited for real-time streaming on resource-constrained hardware. Furthermore, its reliance on purely convolutional operations and reduced parameter count maps naturally to FPGA hardware parallelization.

For the speech separation task, we utilize the causal SuDoRM-RF++ 0.25$\times$ model configuration. The single-channel input is processed by an encoder with 512 basis functions and a kernel size of $T=21$ (2.63 ms at 8 kHz). The separator architecture consists of 4 U-convolutional blocks with 512 input and 256 output channels, utilizing an up-sampling depth of 4. The model estimates masks for two sources, which are element-wise multiplied with the encoded mixture to obtain individual latent representations. These are subsequently reconstructed in the time domain using transposed convolutional layers (see Fig.~\ref{img:AdaptedModel}A). This configuration results in a compact model with approximately 1.63 million parameters and a total size of 6.14 MB.

The separation model is optimized using the negative scale-invariant signal-to-distortion ratio (SI-SDR) loss~\cite{le2019sdr} combined with permutation invariant training (PIT)~\cite{yu2017permutation} to address the label ambiguity problem inherent in multi-speaker separation. Given $N=2$ sources, the permutation invariant loss finds the optimal assignment between estimated and target sources:
\begin{equation}
    \mathcal{L}_{\text{sep}} = \min_{\pi \in \Pi} \left[ -\frac{1}{N} \sum_{i=1}^{N} \text{SI-SDR}(\mathbf{s}_{\pi(i)}, \hat{\mathbf{s}}_i) \right],
\end{equation}
where $\Pi = \{(1,2), (2,1)\}$ represents the two possible permutations for two speakers, and SI-SDR is computed as:
\begin{equation}
    \text{SI-SDR}(\mathbf{s}, \hat{\mathbf{s}}) = 10 \log_{10} \frac{\|\alpha \mathbf{s}\|^2}{\|\alpha \mathbf{s} - \hat{\mathbf{s}}\|^2}, \quad \text{where} \quad \alpha = \frac{\hat{\mathbf{s}}^T \mathbf{s}}{\|\mathbf{s}\|^2}.
\end{equation}
The scaling factor $\alpha$ ensures the metric is invariant to the amplitude of the estimated sources. We employ the Adam optimizer~\cite{kingma2014adam} with an initial learning rate of $10^{-3}$, which is reduced by a factor of 0.3 every 50 epochs. The model is trained with a batch size of 4 on 4-second audio segments randomly sampled from the WSJ0-2mix training set for up to 150 epochs. Early stopping is implemented to halt training when validation SI-SDR fails to improve for 10 consecutive epochs, after epoch 100, to prevent overfitting.

For the speech denoising task, we adapt the architecture with the following modifications to accommodate the different task requirements. The single-channel input is processed by an encoder with 512 basis functions with an increased kernel size of $T=41$ (2.56~ms) following the recommended configuration for 16 kHz in~\cite{tzinis2022compute}, which preserves a similar temporal resolution to the separation task. The separator architecture remains consistent with the separation configuration, featuring 4 U-convolutional blocks with 512 input and 256 output channels. The model estimates a single enhancement mask, which is applied to the encoded mixture before decoding through a single transposed convolutional layer (see Fig.~\ref{img:AdaptedModel}B). This configuration results in a reduced model size of 5.6~MB and approximately 1.48 million parameters. 

The denoising model is optimized using a combination of time-domain and frequency-domain losses following~\cite{defossez2020real}. The objective combines an $L_1$ loss over the raw waveform with a multi-resolution short-time Fourier transform (STFT) loss over spectrogram magnitudes:
\begin{equation}
    \mathcal{L}_{\text{denoise}} = \frac{1}{T} \left[ \|\mathbf{y} - \hat{\mathbf{y}}\|_1 + \sum_{i=1}^{M} L_{\text{stft}}^{(i)}(\mathbf{y}, \hat{\mathbf{y}}) \right],
\end{equation}
where $\mathbf{y}$ and $\hat{\mathbf{y}}$ denote the clean and enhanced signals respectively, $T$ is the signal length, and $M=3$ represents the number of STFT resolutions. Each STFT loss is defined as the sum of spectral convergence ($L_\text{sc}$) and magnitude ($L_\text{mag}$) components:
\begin{equation}
    L_{\text{stft}}(\mathbf{y}, \hat{\mathbf{y}}) = L_\text{sc}(\mathbf{y}, \hat{\mathbf{y}}) + L_\text{mag}(\mathbf{y}, \hat{\mathbf{y}}),
\end{equation}
where the individual components are calculated as:
\begin{align}
    L_\text{sc}(\mathbf{y}, \hat{\mathbf{y}}) &= \frac{\||\text{STFT}(\mathbf{y})| - |\text{STFT}(\hat{\mathbf{y}})|\|_F}{\||\text{STFT}(\mathbf{y})|\|_F},\\
    L_\text{mag}(\mathbf{y}, \hat{\mathbf{y}}) &= \frac{1}{T} \|\log|\text{STFT}(\mathbf{y})| - \log|\text{STFT}(\hat{\mathbf{y}})|\|_1.
\end{align}
Here, $\|\cdot\|_F$ denotes the Frobenius norm and $\|\cdot\|_1$ is the $L_1$ norm. The multi-resolution loss is computed with FFT sizes $\in \{512, 1024, 2048\}$, hop sizes $\in \{50, 120, 240\}$, and window lengths $\in \{240, 600, 1200\}$ samples. We employ the Adam optimizer~\cite{kingma2014adam} with a learning rate of $3 \times 10^{-4}$ and momentums $\beta_1 = 0.9, \beta_2 = 0.999$. The model is trained on the Valentini dataset at a 16 kHz sampling rate for 400 epochs with a batch size of 16. The best model is selected based on its performance on the validation set. 
The training code for both tasks can be found in our~\href{https://github.com/umutcanaltin/audio_task/tree/main}{public repository}.

\subsection{FPGA Platform Specifications}
This study evaluates the proposed speech denoising and separation accelerators on the AMD-Xilinx Kria KV260 Vision AI Starter Kit, which is built around the K26 System-on-Module (SOM) containing a Zynq UltraScale+ MPSoC device. The KV260 is a compact edge-oriented FPGA platform that combines programmable logic with an embedded processing system, making it suitable for low-latency audio inference in embedded and power-constrained settings.

The platform integrates programmable logic resources together with on-board DDR4 memory and ARM processing cores. In our implementation, the programmable logic hosts the synthesized SuDoRM-RF-based inference accelerator, while the ARM Cortex-A53 processing system handles host-side control, model parameter transfer, and runtime interaction with the accelerator. This heterogeneous architecture makes the KV260 well-suited for streaming audio workloads, where lightweight host orchestration is combined with hardware-accelerated neural network execution.

Compared with larger data-center FPGA boards, the KV260 operates under substantially tighter resource and memory constraints. These limitations make it an appropriate target for evaluating whether compact speech denoising and separation models can be deployed efficiently on an edge-class FPGA while still achieving low first-sample latency. In particular, the restricted BRAM, URAM, DSP, and LUT budgets require careful parameter partitioning and selective on-chip caching to reduce the performance penalty of external DDR accesses.

Table~\ref{tab:kv260_specs} summarizes the main hardware characteristics of the KV260 platform used throughout this work.

\begin{table}[h]
\centering
\caption{AMD-Xilinx Kria KV260 platform specifications.}
\label{tab:kv260_specs}
\begin{tabular}{lc}
\toprule
\textbf{Specification} & \textbf{Kria KV260} \\
\midrule
\multicolumn{2}{l}{\textit{Architecture}} \\
Device Family & Zynq UltraScale+ MPSoC \\
Form Factor & SOM + Carrier Card + Thermal Solution \\
Kit Type & Vision AI Kit \\
\midrule
\multicolumn{2}{l}{\textit{Programmable Logic}} \\
System Logic Cells & 256K \\
Block RAM Blocks & 144 \\
UltraRAM Blocks & 64 \\
DSP Slices & 1.2K \\
\midrule
\multicolumn{2}{l}{\textit{Memory Subsystem}} \\
DDR Memory & 4 GB DDR4, non-ECC \\
Primary Boot Memory & 512 Mb QSPI \\
Secondary Boot Memory & SDHC card \\
\midrule
\multicolumn{2}{l}{\textit{Physical Characteristics}} \\
Cooling Solution & Active (Fan + Heatsink) \\
Dimensions & 119 mm $\times$ 140 mm $\times$ 36 mm \\
\bottomrule
\end{tabular}
\end{table}

\subsection{FPGA Implementation}
All FPGA implementations in this work target the AMD-Xilinx Kria KV260 platform. The separation models were manually translated from the trained software model into HLS-synthesizable C++ and implemented as streaming accelerators with AXI4-Stream input/output interfaces and an AXI master interface for parameter access from external DDR memory. Across all versions, the accelerator follows the same high-level computation graph, consisting of an encoder, a bottleneck projection, stacked temporal separation blocks, a mask-generation stage, and an overlap-add decoder. The main differences between implementation versions arise from memory placement, cache organization, and loop-level micro-architectural optimizations, rather than changes to the underlying network structure. 

\paragraph*{Common HLS Design Flow}
All versions were implemented as custom HLS kernels with explicit streaming interfaces. Audio samples are transferred through AXI-stream ports, while model weights are stored in a linear parameter buffer accessed through an AXI master port. In the later versions, the accelerator operates in a two-stage \texttt{mode=0} preload / \texttt{mode=1} inference flow, where selected parameters are first copied from DDR into on-chip memories before streaming inference begins. This separation allows frequently accessed weights to be cached locally, reducing external-memory traffic along the critical path. 

\paragraph*{Speech Separation -- FP32 Implementation (SEP32)}

Speech separation was the first task implemented on the KV260 and drove the bulk of the hardware design exploration in this work. The memory partitioning and buffering decisions developed here were carried directly into the SEP16 and DEN implementations. Three versions were built at FP32 precision, with each iteration targeting the bottleneck exposed by the previous one.

\textbf{SEP32-v1 and SEP32-v2} establish the baseline and expose the primary latency bottleneck. In SEP32-v1, input and output are streamed through AXI4-Stream interfaces while all model parameters are fetched from the external DDR weight buffer at runtime. The temporal state is maintained in shift-register-style history buffers and decoder ring buffers, with aggressively partitioned local arrays for intermediate activations. SEP32-v2 introduces an explicit preload and run execution model, moving the encoder weights, bottleneck weights, and bottleneck biases into on-chip BRAM/URAM before inference begins. The remaining layers still read from DDR, but the front-end feature extraction no longer depends on external memory at runtime, and the original floating-point accumulation order is preserved. The latency gap between the two versions points directly at DDR traffic as the dominant bottleneck rather than compute. This observation guided all subsequent implementation decisions across both separation and denoising.

\textbf{SEP32-v3} extends the on-chip cache beyond the front-end layers. Mask-generation weights, decoder weights, and selected per-block parameters are stored on chip, while the large depthwise convolution kernels remain in DDR due to size constraints. The shift-based encoder history and decoder overlap-add ring buffers are replaced with circular-buffer structures, reducing flip-flop and LUT usage without modifying the network topology. The depthwise-history tensor moves from BRAM to URAM, and the AXI master interface is reconfigured to support longer bursts and more outstanding reads, reducing stall cycles for the remaining DDR accesses. SEP32-v3 achieves the lowest first-sample latency among the FP32 variants and serves as the reference design for all SEP32 results reported in this paper.

\paragraph*{Speech Separation -- F16 Implementation (SEP16)}

Building on the memory partitioning strategy established during the SEP32 exploration, the F16 implementation required fewer design iterations. The caching principles from SEP32-v3 directly informed the starting point for SEP16, reducing the need for exploratory versions. SEP16 implements the same network using end-to-end 16-bit fixed-point arithmetic. The fixed-point data path uses \texttt{ap\_fixed<16,4>} with truncation and wrap-around arithmetic to reduce hardware cost relative to more expensive rounding and saturation modes. Input and output streams are represented as 16-bit AXI words, and the weight buffer holds quantized \texttt{Q4.12}-style parameters in the same linear layout as the FP32 versions.

\textbf{SEP16-v1} is the first resource-tuned 16-bit end-to-end implementation. It adopts the preload/run flow and caches major front-end and back-end parameter groups on chip, including encoder weights, bottleneck weights, mask weights, decoder weights, and selected per-block parameters. The large projection and residual weights are still read from DDR during execution, making this the remaining bottleneck.

\textbf{SEP16-v2} addresses this bottleneck directly. Additional projection and 
residual-path parameters move on chip, including residual matrices, bias vectors, and a preloaded subset of projection weights. Storage placement is refined across URAM, BRAM, and LUTRAM, a more granular placement strategy than used in the FP32 versions. This cuts repeated DDR traffic in the most performance-sensitive matrix-vector stages and is the main reason SEP16-v2 achieves lower first-sample latency than v1. SEP16-v2 is the reference design for all SEP16 results reported in this paper.

\paragraph*{Speech Denoising -- FP32 Implementation (DEN32)}

DEN32 denoising implementation follows the same hardware organization as the separation designs, but is adapted for single-channel output. It uses the same preload/run execution model and replaces shift-register encoder history and decoder overlap-add structures, and circular-buffer organizations. This reduces the flip-flop and LUT overhead without altering the model topology or the original accumulation order. Input and output are streamed through AXI4-Stream interfaces across the full frame sequence.

During preload, the encoder weights, bottleneck weights and biases, mask biases, decoder weights, and selected lightweight per-block parameters are cached on chip. Unlike the final separator designs, the larger projection, residual, and depthwise convolution weights stay in DDR, and only the smaller depthwise biases and activation coefficients are kept locally. At runtime, each temporal block runs projection, multilevel depthwise filtering, feature fusion, and residual accumulation. The depth history tensor resides in URAM to ease BRAM pressure, and the AXI master is set up for longer bursts and more outstanding reads. At the mask generation stage, the mask weight matrix is cached on-chip at reduced precision, while the rest of the datapath remains FP32. This cuts the storage cost in one of the largest cached layers without changing the streaming structure around it.

\paragraph*{Speech Denoising -- F16 Implementation (DEN16)}
DEN16 uses an end-to-end 16-bit fixed-point datapath with \texttt{ap\_fixed<16,4>} arithmetic throughout, using truncation and wrap-around modes over more expensive rounding and saturation. Input and output samples are transported as 16-bit AXI words, and model parameters are stored in a quantized linear weight layout consistent with the FP32 implementation.

The design follows the same preload/run execution model as SEP16, but with a more aggressive on-chip caching strategy than DEN-FP32. During preload, the encoder weights, bottleneck parameters, projection weights and biases, depthwise kernels and biases, residual matrices and biases, mask parameters, and decoder weights are largely moved into on-chip storage. The smaller footprint of the quantized parameter set makes this more feasible than in the FP32 design. Storage placement is refined across URAM, BRAM, and LUTRAM, with larger parameter groups assigned to denser memories and smaller or more frequently accessed structures placed in lower-latency local storage. At the mask-generation layer, the mask matrix is split three ways; one portion in BRAM, a second in LUTRAM, and the remaining tail in DDR fetched on demand. This partitioning provides a compromise between full on-chip residency and memory capacity limits, allowing most mask rows to be served locally while avoiding the storage cost of caching the complete matrix. The runtime otherwise follows the same multiscale streaming organisation as DEN-FP32, with projection, multirate depthwise filtering, feature fusion, residual accumulation, mask estimation, and overlap-add decoding executed in sequence.

\paragraph*{Memory Mapping Strategy}
A central design challenge across all versions is deciding which parameters should be kept on chip and which should remain in external DDR. Because the KV260 has limited BRAM and URAM capacity relative to the size of the full separation model, later implementations use selective caching rather than attempting to embed all weights into on-chip memory. The optimization trajectory across versions follows a consistent pattern: first caching the encoder and bottleneck layers, then moving mask and decoder weights on chip, and finally caching larger projection and residual-path parameters where resource budgets permit. The trade-off is clear: later versions consume more on-chip memory resources but reduce runtime DDR accesses, which in turn lowers first-sample latency.

\paragraph*{Parallelization and Pipelining}
The optimization strategy does not rely on a single global unrolling scheme. Instead, different versions gradually introduce more localized pipelining and partial unrolling in carefully selected loops. Earlier versions prioritize numerical faithfulness and simpler memory access patterns, while later versions increase throughput by buffering the DMA transfer starts and reusing them with partial unrolling in projection and residual computations. At the same time, circular-buffer structures are used to eliminate the overhead of repeatedly shifting temporal histories and decoder states. This combination of selective on-chip caching, localized loop optimization, and state-buffer restructuring forms the core implementation strategy for reducing first-sample latency on the KV260. 

\paragraph*{Runtime Execution on KV260}
On the KV260, the ARM host configures the accelerator, provides the physical address of the parameter buffer, and launches inference through DMA-based streaming transfers. In the preload-enabled versions, the host first invokes the accelerator in preload mode to populate on-chip caches and then switches to inference mode for streaming execution. First-sample latency is measured from the moment DMA transfer is started until the first output sample in the output buffer changes, matching the runtime procedure used in the PYNQ measurement scripts.

\subsubsection{Optimized DEN16 Hardware Architecture}

Building on the memory-mapping trajectory described above, the final optimized DEN16 implementation applies the most aggressive on-chip caching strategy among the evaluated FPGA designs. This subsection therefore focuses specifically on DEN16 rather than all implementation variants. Earlier versions share the same high-level model family, but differ in precision, memory placement, and the amount of parameter reuse captured on chip. DEN16 represents the final hardware organization used for the best latency result, combining 16-bit fixed-point arithmetic, mode-based parameter preload, selective DDR access for low-reuse mask rows, and local URAM/BRAM/LUTRAM storage for high-reuse layers.

\begin{figure}[t]
    \centering
    \includegraphics[width=\columnwidth]{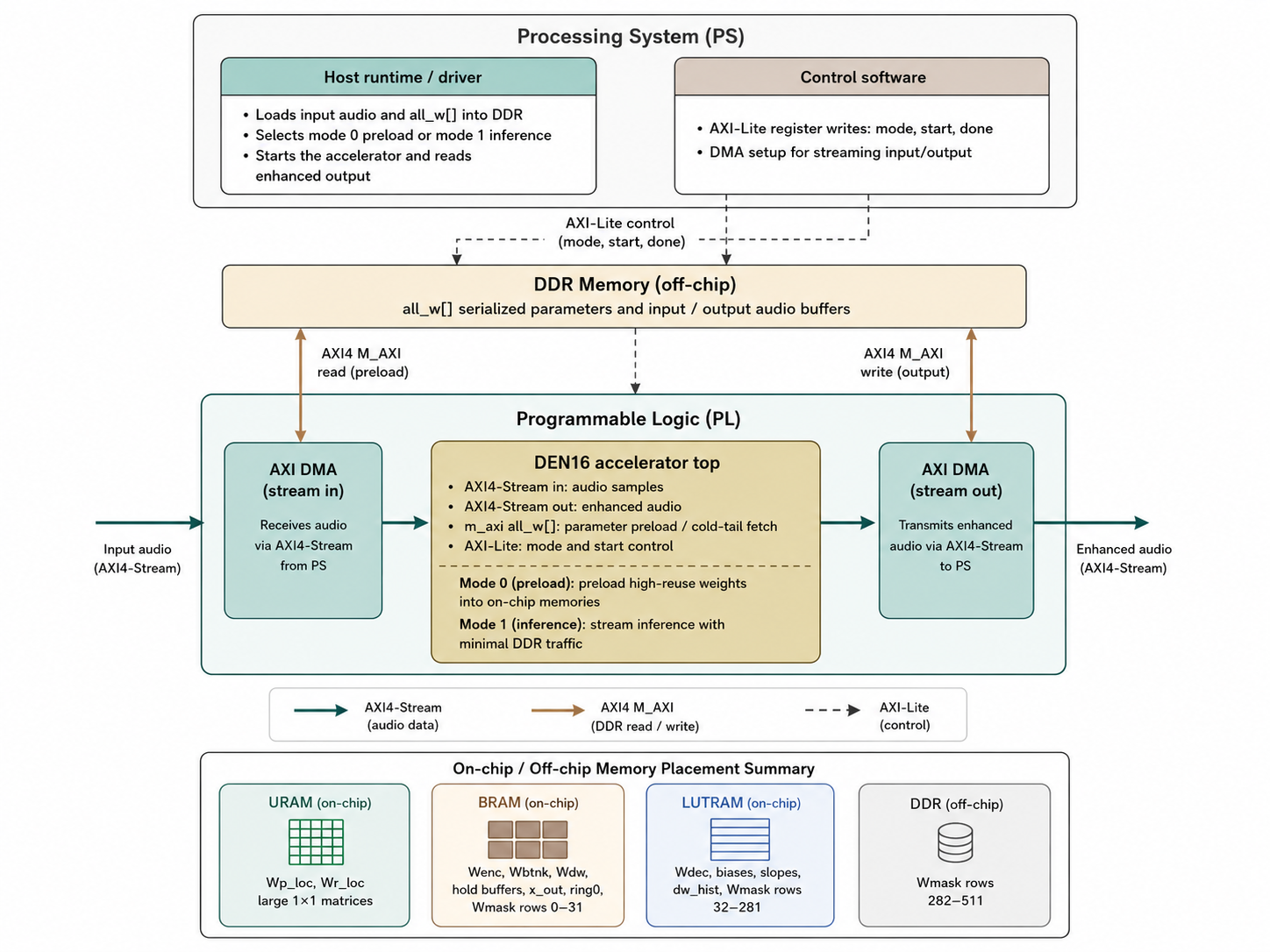}
    \caption{FPGA-SoC execution architecture of the optimized DEN16 accelerator. The processing system manages runtime control, DDR parameter storage, and AXI-stream audio transfers, while the programmable logic hosts the DEN16 accelerator, DMA paths, and on-chip parameter caches. Mode~0 preloads high-reuse parameters into URAM, BRAM, and LUTRAM; mode~1 streams audio through the accelerator while only the low-reuse Wmask tail remains served from DDR.}
    \label{fig:den16_soc}
\end{figure}

Fig.~\ref{fig:den16_soc} separates host-side orchestration from the low-latency streaming datapath implemented in programmable logic. The processing system loads the serialized fixed-point parameter buffer, configures the accelerator through AXI-Lite control registers, and coordinates DMA-based audio movement. During mode~0, high-reuse parameters are preloaded from DDR into local on-chip memories, reducing the amount of external-memory traffic required during inference. During mode~1, audio samples are streamed through the accelerator using AXI4-Stream interfaces, while DDR access is retained only for the low-reuse tail of the mask matrix. This organization reflects the main design principle observed across the implementation variants: reducing runtime DDR access is the primary mechanism for lowering first-sample latency.

\begin{figure}[t]
    \centering
    \includegraphics[width=\columnwidth]{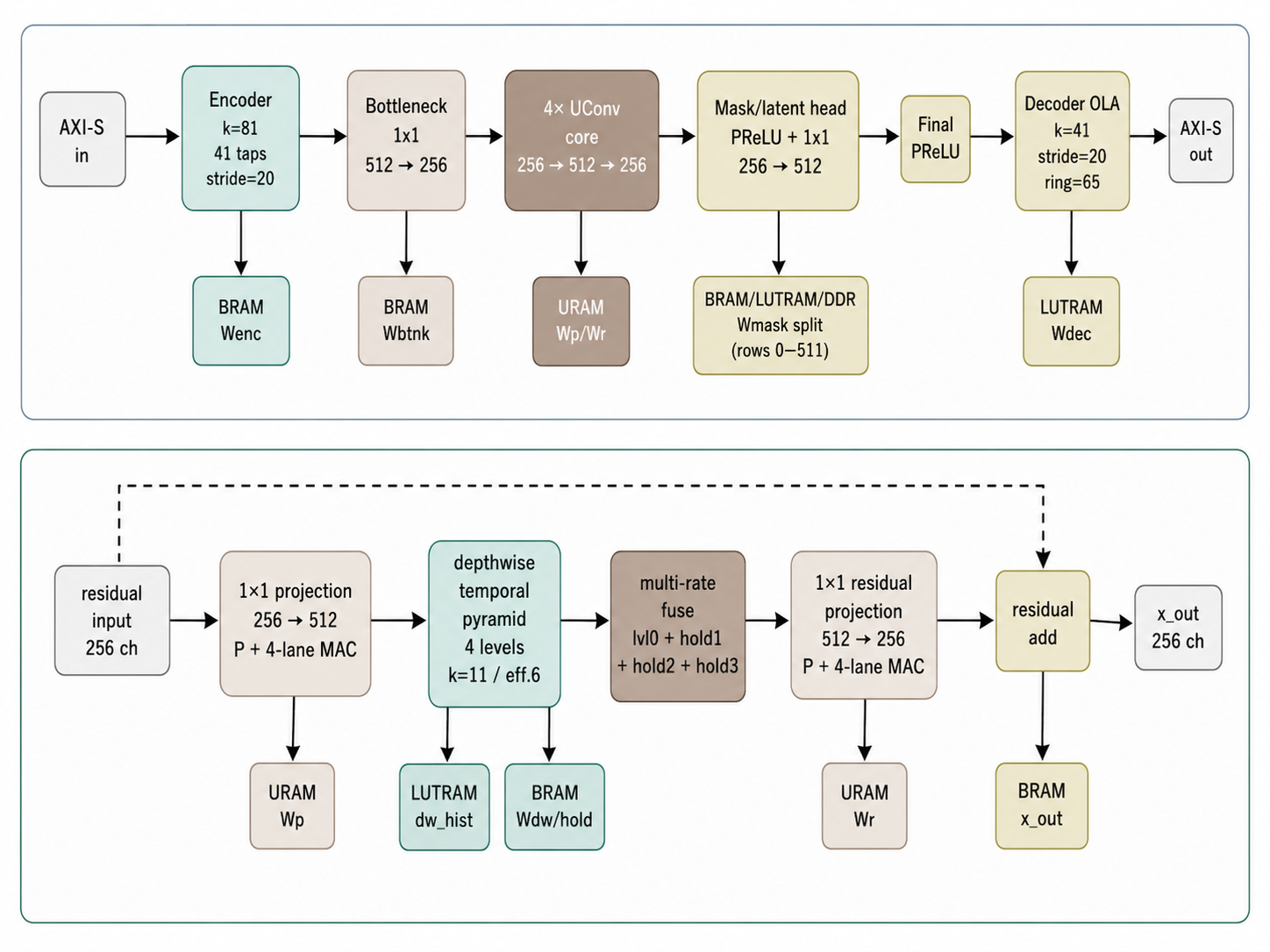}
    \caption{Optimized DEN16 accelerator datapath and UConvBlock microarchitecture. The top-level datapath streams audio through the causal encoder, bottleneck projection, four UConvBlocks, mask/latent head, final PReLU activation, and overlap-add decoder. The lower part expands one repeated UConvBlock, including the 1$\times$1 projection, depthwise temporal pyramid, multi-rate fusion, residual projection, and residual add-back path.}
    \label{fig:den16_datapath}
\end{figure}

The internal DEN16 datapath is shown in Fig.~\ref{fig:den16_datapath}. The accelerator follows the denoising computation graph as a streaming pipeline, with local memory attachments placed near the stages that reuse them most frequently. The UConvBlock detail highlights the repeated residual enhancement unit used inside the core.

Fig.~\ref{fig:den16_datapath} shows how the optimized memory-mapping strategy is reflected inside the accelerator. The encoder, bottleneck, UConv core, mask/latent head, and decoder are arranged as a fixed streaming datapath, while frequently reused parameters are placed in nearby local memories. The 1$\times$1 projection and residual-projection stages use grouped 4-lane MAC computation over locally cached matrices, whereas the depthwise temporal pyramid maintains compact per-channel histories and rate-specific hold buffers. The residual skip path preserves the incoming 256-channel state and adds the learned residual update after the temporal and projection stages. Together, these choices keep the DEN16 datapath suitable for streaming execution while exposing enough local reuse and parallelism to meet the reported first-sample latency target.

\subsection{Experimental Setup}
\label{method:experimental_setup}
To deploy the trained PyTorch models on the KV260, each network was manually translated from the trained software model into HLS-synthesizable C++ using Vitis HLS, with the ARM Cortex-A53 handling host-side control and DMA-based streaming while all neural network computation executes on the programmable logic fabric. The resulting accelerators expose AXI4-Stream input/output interfaces and an AXI master interface for parameter access from external DDR memory. Model weights are exported from PyTorch, quantized where applicable, and serialized into a flat parameter buffer loaded into DDR at runtime.

A primary focus of our deployment study is the impact of precision reduction from 32-bit floating-point (FP32) to 16-bit fixed-point arithmetic (referred to as F16 throughout for brevity), implemented as \texttt{ap\_fixed<16,4>} on the FPGA datapath. This representation uses fixed-point rather than IEEE 754 half-precision floating-point, as fixed-point arithmetic better matches FPGA hardware characteristics by reducing resource usage and improving data locality while maintaining sufficient numerical precision for audio processing. This conversion is motivated by the substantial resource efficiency achievable on FPGA fabric; 16-bit fixed-point operations roughly halve the DSP slice utilization and memory bandwidth requirements compared to FP32, facilitating higher throughput via increased parallelism. We evaluate both tasks, separation and denoising, at both precision levels. The FP32 models serve as our baseline, while the fixed-point variants are obtained through uniform quantization across the entire network, including the encoder, separator U-ConvBlocks, and decoder layers.

The numerical correctness of both precision levels was validated by comparing the hardware outputs with their corresponding PyTorch software reference models. The FP32 designs achieve cosine similarities of 1.0 for both tasks, confirming exact numerical equivalence. The 16-bit fixed-point implementations achieve cosine similarities of 0.986 for speech separation and 0.992 for speech denoising, confirming strong preservation of waveform structure despite reduced precision. These results confirm that the observed deviations are attributable to quantization effects rather than implementation errors, and that the 16-bit designs preserve the underlying waveform structure with cosine similarities above 0.98 across both tasks.

We additionally benchmark both models on an NVIDIA A100 GPU and an Intel Core i7 CPU to provide reference comparisons for the FPGA deployments.

\subsection{Evaluation Metrics}

We assess the quality of enhanced speech signals using established objective metrics. Scale-Invariant Signal-to-Distortion Ratio (SI-SDR)~\cite{le2019sdr} serves as our primary metric for the separation task, reported as SI-SDR improvement (SI-SDRi) in decibels (dB). For the denoising task, we utilize Perceptual Evaluation of Speech Quality (PESQ)~\cite{rix2001perceptual}, standardized as ITU-T P.862.2, which predicts subjective quality on a scale from 0.5 to 4.5. Following established literature in speech enhancement~\cite{defossez2020real}, we also include the composite metrics CSIG, CBAK, and COVL~\cite{hu2007evaluation}, which provide Mean Opinion Score (MOS) estimates of signal distortion, background-noise intrusiveness, and overall perceptual quality, respectively. Short-Time Objective Intelligibility (STOI)~\cite{taal2011algorithm} quantifies intelligibility on a scale of 0 to 1. To address the specific requirements of augmented hearing, we employ the Hearing Aid Speech Perception Index (HASPI)~\cite{kates2014hearing}, which incorporates audiometric models to predict intelligibility for hearing-impaired listeners, and we computed using the PyClarity package~\cite{roa2026cadenza}. PyClarity is configured with an N2 audiogram representing moderate sloping hearing loss. The audiogram is defined at frequencies [250, 500, 1000, 2000, 4000, 6000]~Hz with corresponding hearing threshold levels of [20, 20, 25, 35, 45, 50]~dB~HL, reflecting a common age-related hearing loss profile with greater high-frequency impairment. All quality metrics are computed on the respective test sets, WSJ0-2mix for separation and the Valentini test set for denoising.

For hardware evaluation, we report first-sample latency measured on the KV260 using the PYNQ-based deployment flow. This choice is motivated by the pipelined streaming architecture of the proposed accelerators: in such designs, the most relevant measure for real-time audio processing is the delay until the first valid output sample becomes available, since this directly determines perceived algorithmic latency in a streaming system. After the pipeline is filled, subsequent outputs are produced continuously, making first-sample latency a more meaningful metric than full-frame completion time for low-latency deployment. The Real-Time Factor (RTF) is reported as a secondary metric, computed as the ratio of processing time to one second of audio; an RTF~$<$~1 indicates the capacity for real-time streaming. Hardware cost is reported using post-synthesis utilization estimates from Vitis HLS, covering BRAM18K, DSP, flip-flops (FF), LUTs, and URAM. All implementations target the same KV260 device under the same HLS tool flow, enabling direct comparison of architectural trade-offs across versions. Power consumption is measured using the xmutil platformstats interface on the KV260. The reported CPU latency is averaged over 3,000 inference runs, while the FPGA first-sample latency is the mean of 10 board-level runs; the observed run-to-run variation ranged between 0.08 ms and 0.75 ms across implementations, largely attributable to thermal cycling, which confirms the pipeline behaves consistently under repeated execution.

\section{Experimental Results} 
\label{sec:results}

The evaluation of the proposed system is divided into two primary categories to reflect the constraints of hearing aid deployment. First, Section~\ref{subsec:quality_results} assesses the algorithmic performance of the Speech Separation and Denoising models, focusing on how weight precision affects objective speech quality. Second, the hardware-specific performance, including resource utilization, latency, and power efficiency, is analyzed to determine the feasibility of real-time edge deployment.

\subsection{Speech Enhancement Performance} 
\label{subsec:quality_results}

\paragraph*{Speech Separation}

Table~\ref{tab:separation_results} presents the speech separation results on the WSJ0-2mix test set. The SEP32 model achieved an SI-SDRi of 5.96~dB, with STOI of 0.75 reflecting the speech intelligibility, and PESQ of 2.18 corresponding to fair perceptual quality.

These results fall below the 9.6~dB SI-SDRi reported for the causal C-SuDoRM-RF++ 0.25$\times$ configuration in the original work~\cite{tzinis2022compute}. We attribute this difference to skipping pre-processing steps such as silence removal and static noise suppression that were applied in the original evaluation pipeline. We retain these signal characteristics, as hearing aid applications must process raw acoustic input without the benefit of offline pre-processing. The performance gap reflects the additional challenge of separating speech from unprocessed mixtures containing low-level recording artifacts found in the WSJ0 corpus.

Conversion to float point 16 (SEP16) reduced model size by 50\%, from 6.14~MB to 3.07~MB. The SEP16 model marginally outperforms the SEP32 model, achieving an SI-SDRi of 6.03~dB while maintaining equivalent STOI and PESQ scores. This indicates that the reduced precision does not degrade the learned representations and is also better suited for implementation on smaller FPGA chips such as the KV260 accelerator.

\begin{table}[!ht]
\centering
\small
\caption{Speech Separation Results on WSJ0-2mix Test Set.}
\label{tab:separation_results}
\begin{tabular}{lcccc}
\toprule
\textbf{Model} & \textbf{SI-SDRi (dB)} & \textbf{STOI} & \textbf{PESQ} & \textbf{Size (MB)} \\
\midrule
Unprocessed & 0.00 & 0.66 & 1.81 & — \\
SEP32 & 5.96 & 0.75 & 2.18 & 6.14 \\
SEP16 & 6.03 & 0.75 & 2.17 & 3.07 \\
\bottomrule
\end{tabular}
\end{table}

\paragraph*{Speech Denoising}

Table~\ref{tab:denoising_results} presents the speech denoising results on the Valentini test set. The DEN32 model achieved a PESQ of 2.41 and STOI of 0.93, representing improvements of 0.44 and 0.01 over the unprocessed baseline, respectively. The composite metrics indicate preserved signal quality (CSIG: 4.01), effective background noise suppression (CBAK: 3.23), and improved overall quality (COVL: 3.35).

We adopt the training procedure and loss functions from Defossez et al.~\cite{defossez2020real}, combining L1 waveform loss with multi-resolution STFT loss. However, we replace their DEMUCS architecture with the adapted SuDoRM-RF++ model to reduce model size for edge deployment. The causal DEMUCS model achieves higher scores (PESQ: 2.91, STOI: 0.95) but requires approximately 135~MB, compared to 5.6~MB for our approach, a 24$\times$ reduction in model size. Notably, background noise suppression performance (CBAK: 3.23 vs 3.25) remains comparable despite this difference, suggesting the architecture efficiently captures the features necessary for noise attenuation.

For hearing aid applications, we additionally report the Hearing Aid Speech Perception Index (HASPI), computed using an N2 audiogram representing moderate sloping hearing loss. The DEN32 model achieved a HASPI score of 0.90, indicating that speech intelligibility cues relevant to hearing-impaired listeners are preserved through the enhancement process.

Conversion to float point 16 (DEN16) precision reduced model size to 2.81~MB while maintaining identical performance across all metrics. This consistency between precision levels, combined with the 50\% memory reduction, supports the feasibility of deploying the denoising model on resource-constrained hardware without sacrificing enhancement quality.

\begin{table}[!ht]
\centering
\small
\caption{Speech denoising results on the Valentini test set.}
\label{tab:denoising_results}
\resizebox{\columnwidth}{!}{%
\begin{tabular}{lccccccc}
\toprule
\textbf{Model} & \textbf{PESQ} & \textbf{STOI} & \textbf{CBAK} & \textbf{COVL} & \textbf{CSIG} & \textbf{HASPI} & \textbf{Size (MB)} \\
\midrule
Unprocessed & 1.97 & 0.92 & 2.80 & 2.70 & 3.32 & 0.89 & — \\
DEN32  & 2.41 & 0.93 & 3.23 & 3.35 & 4.01 & 0.90 & 5.60 \\
DEN16  & 2.41 & 0.93 & 3.23 & 3.35 & 4.01 & 0.90 & 2.81 \\
\bottomrule
\end{tabular}%
}
\end{table}

\subsection{FPGA Results: Speech Separation}
\label{subsec:sep_fpga}

\begin{table}[t]
\centering
\small
\setlength{\tabcolsep}{3.5pt}
\caption{FPGA results for speech separation and denoising implementations on the Kria KV260.}
\label{tab:fpga_results_all}
\resizebox{\columnwidth}{!}{%
\begin{tabular}{llrrrrrr}
\toprule
\textbf{Task} & \textbf{Model} & \textbf{First-sample} & \textbf{BRAM} & \textbf{DSP} & \textbf{FF} 
    & \textbf{LUT} & \textbf{URAM} \\
 & & \textbf{latency (ms)} & \textbf{18K} & & & & \\
\midrule
\multirow{5}{*}{Speech Separation}
 & SEP32-v1 & 157.6 & 119 & 13  & 53,278  & 43,196 & 0  \\
 & SEP32-v2 & 114.0 & 141 & 109 & 60,255  & 50,158 & 32 \\
 & SEP32-v3 & 44.0  & 254 & 109 & 100,109 & 81,599 & 56 \\
 & SEP16-v1 & 27.3  & 138 & 821 & 72,205  & 65,697 & 48 \\
 & SEP16-v2 & 16.0  & 263 & 875 & 80,102  & 57,299 & 58 \\
\midrule
\multirow{2}{*}{Speech Denoising}
 & DEN32 & 41.2 & 238 & 210 & 111,045 & 91,737 & 47 \\
 & DEN16 & 9.7 & 253 & 899 & 83,060  & 90,863 & 64 \\
\bottomrule
\end{tabular}%
}
\end{table}

All our implementations target the same KV260 device and were synthesized under the same Vitis HLS tool flow, enabling direct comparison of architectural trade-offs across versions. The first-sample latency is reported as the primary runtime metric, and the hardware cost is drawn from post-synthesis utilization estimates covering BRAM18K, DSP, flip-flops, LUTs, and URAM.

\paragraph*{SEP32 Results}

Among the SEP32 implementations (see Table~\ref{tab:fpga_results_all}), the baseline SEP32-v1 exhibits the highest first-sample latency at 157.6~ms while using 119 BRAM18K, 13 DSP, 53,278 FF, and 43,196 LUTs. Introducing on-chip caching in SEP32-v2 reduces first-sample latency to 114.0~ms, at the cost of increased on-chip memory and DSP usage (141 BRAM18K, 109 DSP, 60,255 FF, 50,158 LUTs, and 32 URAM). A substantially more aggressive caching and memory-partitioning strategy is used in SEP32-v3, which lowers first-sample latency to 44.0~ms while increasing utilization to 254 BRAM18K, 109 DSP, 100,109 FF, 81,599 LUTs, and 56 URAM. SEP32-v3 is the SEP32 reference design in all subsequent comparisons.

\paragraph*{SEP16 Results}

Building on the memory partitioning strategy from SEP32-v3, the SEP16 implementation required fewer design iterations and achieves substantially lower first-sample latency than the SEP32 variants, dropping from 44.0~ms to 16.0~ms at the optimized configuration (see Table~\ref{tab:fpga_results_all}).

SEP16-v1 reaches 27.3~ms with a resource footprint of 138 BRAM18K, 821 DSP, 72,205 FF, 65,697 LUTs, and 48 URAM. The final optimized fixed-point design, SEP16-v2, further reduces first-sample latency to 16.0~ms while increasing on-chip memory usage to 263 BRAM18K and 58 URAM, together with 875 DSP, 80,102 FF, and 57,299 LUTs. The lower latency of SEP16-v2 relative to v1 comes from storing more projection and residual-path parameters on chip, cutting external DDR accesses during inference. The fixed-point design space gives the best latency on the KV260, but only when sufficient BRAM and URAM are committed to local parameter caching.

\paragraph*{Overall Trends}

Across all separation implementations, the main latency reduction comes from progressively increasing the amount of model state and parameters cached on chip. The early SEP32 versions rely heavily on runtime DDR accesses and pay for it in latency. Later SEP32 variants and especially the SEP16 designs shift more weights and intermediate state into BRAM and URAM, cutting memory-access overhead along the critical inference path. The fastest separation result is SEP16-v2 at 16.0~ms, and the slowest is SEP32-v1 at 157.6~ms. On the KV260, low-latency streaming inference is primarily limited by memory placement and data movement costs, rather than raw arithmetic throughput.

\subsection{FPGA Results: Speech Denoising}
\label{subsec:enh_fpga}

The denoising accelerator targets the same KV260 platform and follows the same 
HLS-based deployment flow described in Section~\ref{subsec:sep_fpga}. Unlike the separation task, the denoising model's single-output architecture reduces memory pressure sufficiently that only one FP32 and one F16 implementation were needed to achieve the best achievable latency at each precision level.

\paragraph*{DEN32 Results}

DEN32 achieves a first-sample latency of 41.2~ms, with a post-synthesis resource footprint of 238 BRAM18K, 210 DSP, 111,045 FF, 91,737 LUTs, and 47 URAM (see Table \ref{tab:fpga_results_all}). Relative to the earlier SEP32 variants, this design already incorporates a more aggressive optimization strategy, including circular-buffer-based history management and selective on-chip caching of encoder, bottleneck, mask, and decoder parameters, while leaving some large tensors to be fetched from external memory at runtime. The latency is substantially lower than the SEP32 baselines, though it still reflects the overhead of float32 arithmetic and partial reliance on off-chip accesses in the critical path.

\paragraph*{DEN16 Results}

DEN16 reaches the lowest latency among the enhancement variants, with a first-sample latency of 9.7~ms (see Table~\ref{tab:fpga_results_all}). This design uses 253 BRAM18K, 899 DSP, 83,060 FF, 90,863 LUTs, and 64 URAM. Compared with DEN32, the fixed-point design reduces runtime latency by caching a larger fraction of the model on chip and replacing float32 
arithmetic with lower-precision fixed-point computation. The design makes extensive use of BRAM and fully occupies the available URAM on the KV260, indicating that the latency gain comes primarily from more aggressive on-chip parameter placement and reduced dependence on external DDR during inference. Notably, DEN16 is the only implementation across both tasks to fall within the 10~ms clinical latency threshold for hearing prosthetics.

\paragraph*{Overall Trends}

The denoising results follow the same pattern seen in separation: first-sample latency tracks closely with on-chip data locality rather than arithmetic throughput. DEN32 already benefits from circular history buffers and selective caching, which put it well ahead of the slower SEP32 baselines. DEN16 goes further, moving more model state on chip and switching to fixed-point arithmetic that maps more efficiently onto the KV260 fabric. The result is a $4.23\times$ latency reduction from DEN32 to DEN16, dropping from 41.2~ms to 9.7~ms, which is a larger gain than the corresponding FP32-to-F16 reduction in separation. For low-latency streaming on the KV260, memory placement and precision selection are the decisions that matter most.

\subsection{Latency and Clinical Feasibility}
\label{subsec:latency_results}

\begin{figure}[!ht]
    \centering
    \includegraphics[width=0.9\linewidth]{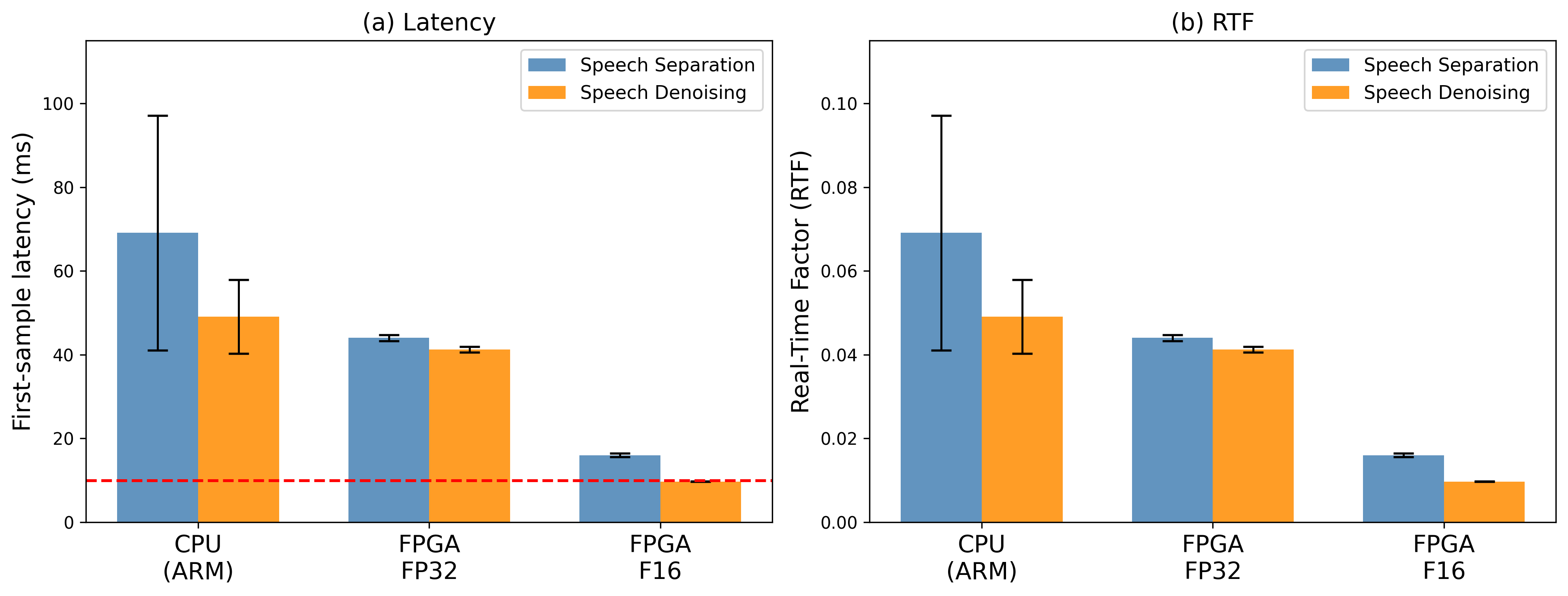}
    \caption{First-sample latency (a) and equivalent Real-Time Factor (b) for the best-performing FP32 and F16 implementations on the Kria KV260, compared against the ARM CPU baseline. The dashed red line marks the 10~ms clinical latency threshold for hearing prosthetics~\cite{stone2003tolerable}. The error bars represent the standard deviation. DEN16 is the only configuration to fall within the clinical threshold.}
    \label{img:latency_comparison}
\end{figure}

In Figure~\ref{img:latency_comparison} we show the first-sample latency and RTF for the CPU baseline and the reference FPGA designs across both tasks, measured over one second of audio. Only the best-performing implementation at each precision level is included: SEP32-v3, SEP16-v2, DEN32, and DEN16.

For speech separation, the ARM CPU baseline processes one second of audio in 69.1~ms (RTF: 0.069). SEP32 on the FPGA reduces this to 44.0~ms (RTF: 0.044), a $1.57\times$ reduction on the same embedded platform. SEP16 further reduces latency to 16.0~ms (RTF: 0.016), a $2.76\times$ improvement over SEP32 and a $4.33\times$ improvement over the CPU baseline. While neither configuration meets the 10~ms clinical threshold for hearing prosthetics, SEP16 narrows the gap considerably.

For speech denoising, the CPU baseline processes one second of audio in 49.1~ms (RTF: 0.049). DEN32 reduces this to 41.2~ms (RTF: 0.041), a modest improvement that still reflects the overhead of float32 arithmetic and partial DDR dependence. DEN16 achieves a first-sample latency of 9.7~ms (RTF: 0.010), which falls within the 10~ms clinical threshold~\cite{stone2003tolerable}. The $4.23\times$ reduction from DEN32 to DEN16 is the largest precision-driven latency gain we observed across all configurations evaluated and exceeds the corresponding gain in separation ($2.76\times$), reflecting the lower memory pressure of the single-output denoising architecture.

For reference, the A100 GPU achieves first-sample latencies of 2.6~ms and 3.1~ms for SEP32 and SEP16, respectively, and 3.1~ms and 3.4~ms for DEN32 and DEN16, all well within the clinical threshold. All four configurations comfortably meet the clinical threshold, confirming that the models can achieve real-time performance when available compute is unconstrained.

Table~\ref{tab:literature_comparison} places the best results from this work in the context of prior FPGA-based speech denoising work. Flamis et al.~\cite{flamis2022fpga} deployed an SE-FCN on the XC7Z020 using INT8 quantization via Vitis AI, demonstrating DNN-based denoising on a resource-constrained SoC-FPGA, but do not report latency or RTF, making real-time feasibility difficult to assess. Ni et al.~\cite{ni2025fpga} deployed an LSTM-based denoising model on the AMD RFSoC4x2 and reported a total processing latency of 7,500~ms with an RTF of 1.875, meaning the system cannot process audio in real time. In contrast, SEP16 achieves a first-sample latency of 16.0~ms for speech separation and DEN16 achieves 9.7~ms for denoising, both on a more resource-constrained platform and with higher absolute PESQ scores. To our knowledge, DEN16 is the first fully DNN-based time-domain speech enhancement implementation on an embedded FPGA that has been reported to meet the 10~ms clinical latency threshold for hearing prosthetics.

\begin{table*}[ht]
\centering
\caption{Comparison with prior FPGA-based speech denoising work. Resource utilization is reported as absolute counts.
$^{*}$Ni et al.\ report total processing latency for a 4-second utterance; first-sample latency is not reported separately. $^{\dagger}$Flamis et al.\ report PESQ improvement over a non-DNN baseline
rather than an absolute score.}
\label{tab:literature_comparison}
\small
\begin{tabular}{lllccccccc}
\toprule
\textbf{Reference} & \textbf{Platform} & \textbf{Model}
    & \textbf{Prec.} & \textbf{BRAM} & \textbf{DSP} & \textbf{LUT}
    & \textbf{Latency (ms)} & \textbf{RTF} & \textbf{PESQ} \\
\midrule
Flamis et al.~\cite{flamis2022fpga}
    & XC7Z020  & SE-FCN
    & INT8 & 118 & 194 & 30,074
    & ---           & ---   & $+0.5^{\dagger}$ \\
Ni et al.~\cite{ni2025fpga}
    & RFSoC4x2 & LSTM
    & FP32 & 21  & 459 & 61,624
    & 7,500$^{*}$   & 1.875 & 1.897 \\
\midrule
\textbf{This work (DEN16)}
    & KV260    & SuDoRM-RF++
    & F16 & 253 & 899 & 90,863
    & \textbf{9.7} & \textbf{0.010} & \textbf{2.41} \\
\bottomrule
\end{tabular}
\end{table*}

\subsection{Power Efficiency and Trade-offs}
We evaluated power consumption for the best performing SEP and DEN FPGA implementations using the Vivado power report derived from the post-place-and-route netlist (see Table~\ref{tab:power_results}). Since the proposed accelerators operate as pipelined streaming architectures, we report energy-to-first-sample rather than total energy per frame, as the first valid output sample determines the effective system response time. The corresponding energy-to-first-sample is computed as:
\begin{equation}
E_{\text{first}} = P_{\text{on-chip}} \times T_{\text{first-sample}}.
\end{equation}

Among the evaluated designs, SEP32 has an estimated on-chip power of 3.995~W, SEP16 3.707~W, DEN32 3.905~W, and DEN16 4.199~W. Although these values span a narrow range, the large latency differences produce substantially different energy-to-first-sample figures: 175.9~mJ for SEP32, 59.2~mJ for SEP16, 160.9~mJ for DEN32, and 40.9~mJ for DEN16, corresponding to $3.0\times$ and $3.9\times$ energy reductions for the fixed-point separation and denoising designs, respectively.

\begin{table*}[t]
\centering
\small
\caption{Estimated on-chip power and energy-to-first-sample for representative KV260 implementations.}
\label{tab:power_results}
\begin{tabular}{lrrrr}
\toprule
\textbf{Model} & \textbf{First-sample latency (ms)} & \textbf{On-chip power (W)} & \textbf{Energy-to-first-sample (mJ)} & \textbf{PS share (\%)} \\
\midrule
SEP32 & 44.0 & 3.995 & 175.9 & 60.6 \\
SEP16 & 16.0 & 3.707 & 59.2  & 65.3 \\
DEN32 & 41.2 & 3.905 & 160.9 & 61.9 \\
\textbf{DEN16} & \textbf{9.7} & \textbf{4.199} & \textbf{40.9} & \textbf{57.6} \\
\bottomrule
\end{tabular}
\end{table*}

The processing system (PS) remains approximately constant at 2.419~W across all designs, accounting for 57.6\% -- 65.3\% of total on-chip power, which means PL datapath changes have a comparatively modest impact on total power. The floating-point designs spend more power in clock, signal, and logic networks due to wider datapaths and larger control structures, while the fixed-point designs shift more computation into DSP-heavy arithmetic and aggressively cache on-chip memories. DEN16 illustrates this trade-off directly: despite having the highest total on-chip power among the four designs, it achieves the lowest energy-to-first-sample because it is also the fastest implementation.

Overall, these results show that latency optimization and power optimization are not identical objectives. For the KV260, the fixed-point SEP16 and DEN16 operating points offer the most favorable balance among latency, resource utilization, and energy-to-first-sample. The reported power values are based on implemented-netlist estimation with vectorless activity assumptions and should be interpreted as comparative design estimates rather than exact board-level measurements.

\section{Discussion}
\label{sec:discussion}

In this study, we deployed time-domain speech enhancement models on an embedded FPGA platform to characterize the feasibility gap between current hardware capabilities and the latency constraints of hearing aids. By implementing both speech separation and denoising using the lightweight SuDoRM-RF++ architecture on the Xilinx Kria KV260, we characterized the computational demands, memory footprint, and power consumption inherent to each task across FP32 and 16-bit fixed-point precision. Our results demonstrate that reducing precision halves the model memory footprint without compromising objective speech quality, and that aggressive on-chip parameter caching is the primary lever for latency reduction on this platform. DEN16 achieves a first-sample latency of 9.7~ms, meeting the 10~ms clinical threshold for hearing prosthetics~\cite{stone2003tolerable} and establishing that DNN-based time-domain denoising is feasible on current embedded FPGA hardware. While speech separation remains just outside this threshold and the platform power envelope far exceeds hearing aid budgets, these findings provide a way to characterize the resource requirements for embedded DNN-based speech enhancement.

The selection of the time-domain SuDoRM-RF++ architecture addresses both the latency requirements and hardware constraints of hearing aids. Frequency-domain models incur inherent algorithmic latency from STFT windowing; a typical configuration with a 32~ms window and 16~ms hop introduces approximately 48~ms delay before neural processing begins~\cite{wang2018supervised}. Time-domain convolutional networks bypass this constraint by operating directly on the raw waveform with frame sizes as low as 2-4~ms~\cite{luo2019conv,tzinis2022compute}. The purely convolutional structure of SuDoRM-RF++ also maps efficiently to FPGA DSP slices, unlike recurrent~\cite{luo2020dual} or attention-based~\cite{subakan2021attention} architectures that achieve higher separation quality but require sequential processing incompatible with pipelined streaming inference. Choosing a compact SuDoRM-RF++ does limit separation performance compared to larger architectures such as Causal SepFormer~\cite{subakan2021attention} and DPRNN~\cite{luo2020dual}, which require $26$--$130\times$ more parameters and exceed the KV260 memory budget. This limitation is compounded by the inherent difficulty of the separation task: it must resolve two sources that share similar spectro-temporal characteristics and are subject to permutation ambiguity during training~\cite{yu2017permutation}, whereas denoising exploits statistical speech-noise differences that compact filterbanks capture more readily. Rather than deeper architectures, richer input representations offer a more feasible path forward, as binaural spatial cues have been shown to improve separation without additional parameters~\cite{olalere2025leveraging}, and multi-microphone beamforming can deliver cleaner inputs to compact neural backends~\cite{heymann2016neural}.

While architectural choices address computational complexity, memory bandwidth remains a critical constraint for embedded FPGA deployment. Across all implementations, latency tracks directly with the fraction of model parameters cached in BRAM and URAM rather than with DSP utilization, confirming that the deployment challenge is fundamentally one of data movement rather than computation only. Reducing numerical precision is a common strategy for addressing this in resource-constrained FPGA deployments, with most accelerator studies targeting INT8 or lower precision combined with pruning and quantization-aware training~\cite{guo2019dl}. For convolutional architectures, the benefits of aggressive quantization are more limited than for attention or fully-connected networks, as compute cost is dominated by convolution operations rather than memory access~\cite{wu2020integer}. Furthermore, audio enhancement filter weights are particularly sensitive to quantization noise since they encode fine spectral distinctions that coarse representations can blur. We therefore evaluated 16-bit precision as a practical starting point, as it halves memory bandwidth and storage requirements while maintaining sufficient dynamic range for audio signals. On both tasks, the 16-bit model matched or marginally exceeded FP32 performance, confirming that the architecture tolerates precision reduction without retraining. The hardware datapath uses Q4.12 fixed-point arithmetic rather than IEEE F16 floating-point. However, the numerical validation against the software reference confirms cosine similarities of 0.986 and 0.992 for separation and denoising, respectively, indicating that precision-induced deviations are small and attributable to quantization effects. Further compression via pruning or INT8 quantization can build on this baseline, with the 16-bit results providing a reference for acceptable quality degradation~\cite{tan2021towards}.

The most significant gap between this work and wearable deployment is power consumption. Hearing aids must operate within a few-milliwatt budget to achieve acceptable battery life~\cite{gerlach2022survey,kates2008hearing}, and application-specific hearing-aid processors have demonstrated that this is achievable in silicon.~\cite{kim2019high} report a 65~nm ASIP-based hearing aid SoC consuming 1.3~mW at 1~V and 8~MHz. By contrast, the KV260 draws 3--4~W in these experiments, with the processing system alone consuming 2.419~W, exceeding the entire hearing aid power budget by nearly three orders of magnitude. This gap reflects the nature of the KV260 as a characterization platform rather than a deployment target, and the value of these results lies in quantifying the on-chip memory and latency trade-offs that future low-power implementations must satisfy. Beyond power, the evaluation is single-channel only and can be extended to a multi-channel model. Improving the performance of smaller models for speech separation while reducing latency to meet the clinical threshold are both directions for future work. Nevertheless, the results establish a concrete characterization of the resource requirements for embedded DNN-based speech enhancement and demonstrate that the clinical latency threshold is reachable on current embedded FPGA hardware for the denoising task.

\section{Conclusion}
\label{conclusion}
This work presented a feasibility characterization of time-domain DNN-based speech enhancement on the AMD-Xilinx Kria KV260 embedded FPGA, targeting the latency and resource constraints of hearing aids. Both speech separation and denoising were implemented using the SuDoRM-RF++ architecture at FP32 and 16-bit fixed-point precisions, with systematic hardware design iterations revealing on-chip parameter caching as the primary lever for latency reduction. The optimized fixed-point denoising accelerator, DEN16, achieves a first-sample latency of 9.7~ms and meets the 10~ms clinical threshold for hearing prosthetics, while the separation accelerator reaches 16.0~ms. Precision reduction from FP32 to 16-bit fixed-point preserved or marginally improved objective speech quality across all metrics, with the halved memory footprint enabling more aggressive on-chip caching in the fixed-point designs. The platform power consumption of 3--4~W remains above hearing aid budgets, confirming that the KV260 serves as a characterization substrate rather than a deployment target. Future work will investigate structured pruning and quantization-aware training to further reduce memory pressure while retaining objective performance. Also, the use of multi-microphone input to improve separation performance, and custom low-power ASIP or ASIC implementations targeting the on-chip memory and latency trade-offs quantified in this study.

\section{Acknowledgements}
\noindent This work is part of the INTENSE consortium, which has received funding from the NWO Cross-over Grant No. 17619. 

\bibliographystyle{ieeetr}
\bibliography{Main}

\end{document}